\documentstyle[12pt,fleqn]{article}  
\begin{document}  
\begin{titlepage}  
  
\begin{center}  
\large\bf On the manifold of the Laughlin problem unique solutions  
\end{center}  
\vspace{2.5cm}  
\renewcommand{\thefootnote}{\fnsymbol{footnote}}  
\begin{center}  
B.A. Lysov, O.F. Dorofeyev \\  
{\sl Physical Faculty, Moscow State University,\\  
119899, Moscow, Russia}  
\end{center}  
  
\bigskip  
  
\bigskip  
  
\begin{abstract}  
Solutions, exactly expressed in terms of elementary functions  
(unique Laughlin states), of the  correlated motion problem for  
a pair of  2D-electrons in a constant and uniform magnetic  
field have been shown to exist for a certain relation between  
the magnetic field induction and the electron charge.  
Arguments that can help to understand the physical meaning of these  
remarkable magnetic field values  have been provided.  
The special interest to this problem  is justified by the importance  
of the  new state of matter recently observed.  
  
\bigskip  
  
\bigskip  
  
\bigskip  
  
\bigskip  
  
\bigskip

E-mail:dorofeyev@ofme.phys.msu.su  
  
\end{abstract}  

\vspace{0.3cm}  
\vfill  
\end{titlepage}  
  
Let us consider the motion of a pair of 2D-electrons in a constant  
and uniform magnetic field. In this problem, the motion of the electrons  
center  
of mass can be considered separately from their relative motion.  
  
The stationary Schr\"odinger equation for the relative motion  
(the Laughlin problem) has the form  
\begin{equation}  
\left\{ \frac{1}{m^{\ast }}\left[ \left( \widehat{p}_{x}-\frac{eB}{4c}%
\widehat{y}\right) ^{2}-\left( \widehat{p}_{y}+\frac{eB}{4c}\widehat{x}%
\right) ^{2}\right] +\frac{e^{2}}{\sqrt{x^{2}+y^{2}}}\right\} \Psi =E\Psi \;.  
\label{1}  
\end{equation}  
\bigskip  
Here $m^{\ast }$ is the effective electron mass in the given  
heterostructure, $e$ is the absolute value of the electron charge, $B$ is the magnetic  
field induction. The symmetric gauge for the vector potential in the  
above equation has been used  
\begin{equation}  
\overrightarrow{A}=-\frac{B}{2}y\overrightarrow{e}_{x}+\frac{B}{2}x%
\overrightarrow{e}_{y}=\frac{B}{2}r\overrightarrow{e}_{\varphi }\;.  
\label{2}  
\end{equation}  
Equation (1) formally corresponds to the Schr\"odinger equation  
for a mass $m^{\ast }/2$ and charge $-e/2$ particle interacting with the  
uniform  
magnetic field (2) and with a fixed charge $-2e$ placed at the origin.

Equation (1) is conveniently transformed to dimensionless variables.  
To this end, we introduce the magnetic length $l_{B}=\sqrt{2\hbar c/\left(  
eB\right) }$ and the dimensionless energy eigenvalue $\lambda =2E/\left(  
\hbar \omega _{c}\right) ,$ where $\omega _{c}=eB/\left( m^{\ast }c\right)$ is  
the cyclotron frequency. Upon introduction of the new dimensional variables  
\begin{equation}  
x=l_{B}\xi ,\qquad y=l_{B}\eta ,\qquad \rho =\sqrt{\xi ^{2}+\eta ^{2}},  
\label{3}  
\end{equation}  
equation (1) takes the form  
  
\begin{equation}  
\left[ \left( \widehat{p}_{\xi }-\frac{1}{2}\widehat{\eta }\right)  
^{2}+\left( \widehat{p}_{\eta }+\frac{1}{2}\widehat{\xi }\right) ^{2}+\frac{a%
}{\rho }\right] \Psi =\lambda \Psi \;.  
\label{4}  
\end{equation}  
\bigskip  
Here $a=\sqrt{B_{0}/B}$ and $\ B_{0}=2cm^{\ast 2}e^{3}/\hbar ^{3}\simeq  
4.7\times 10^{9}\left( m^{\ast }/m\right) ^{2}$ is the critical magnetic  
 field. In the polar coordinates, equation (4) reads  
  
\begin{equation}  
\left[ -\left( \frac{\partial ^{2}}{\partial \rho ^{2}}+\frac{1}{\rho }\frac{%
\partial }{\partial \rho }\right) -\frac{1}{\rho ^{2}}\frac{\partial ^{2}}{%
\partial \varphi ^{2}}-i\frac{\partial }{\partial \varphi }+\frac{1}{4}\rho  
^{2}+\frac{a}{\rho }\right] \Psi =\lambda \Psi .  
\label{5}  
\end{equation}  
\bigskip  
The wave function is looked for in the form  
  
\begin{equation}  
\Psi =\exp \left( il\varphi \right) R\left( \rho \right) .  
\label{6}  
\end{equation}  
 The radial part of the wave function is described by the  equation  
  
\begin{equation}  
\left[ -\left( \frac{\partial ^{2}}{\partial \rho ^{2}}+\frac{1}{\rho }\frac{%
\partial }{\partial \rho }\right) +\frac{l^{2}}{\rho ^{2}}+l+\frac{1}{4}\rho  
^{2}+\frac{a}{\rho }\right] R\left( \rho \right) =\lambda R\left( \rho\right) .  
\label{7}  
\end{equation}  
\bigskip  
Here \ $l=0\pm 1,\pm 2,...$ are  eigenvalues of the operator  
\ $L_{z}=-i\frac{\partial }{\partial\varphi }$.  
  
In the following, only the case $l=0$\ \ is considered, and hence  
the equation takes the form  
  
\begin{equation}  
\left[ -\left( \frac{\partial ^{2}}{\partial \rho ^{2}}+\frac{1}{\rho }\frac{%
\partial }{\partial \rho }\right) +\frac{1}{4}\rho ^{2}+\frac{a}{\rho }%
\right] R\left( \rho \right) =\lambda R\left( \rho \right) .  
\label{8}  
\end{equation}  
\bigskip  
The Hamiltonian operator in the left-hand side of (8)  
can be written in the form  
  
\begin{equation}  
\widehat{H}=\widehat{p}_{\rho }^{2}+\frac{1}{4}\rho ^{2}+\frac{1}{4\rho ^{2}}%
+\frac{a}{\rho }\;.  
\label{9}  
\end{equation}  
\bigskip  
Here  
  
\begin{equation}  
\widehat{p}_{\rho }=\frac{1}{i}\left( \frac{\partial }{\partial \rho }+\frac{%
1}{2\rho }\right) .  
\label{10}  
\end{equation}  
\bigskip  
is the radial momentum operator,  selfadjoint in the Hilbert space $L^{2}\left(  
0,\infty ,\rho d\rho \right) .$ It should be mentioned that the term  
$1/\left( 4\rho ^{2}\right) $\ \ has appeared in (9).  
  
Radial eigenfunctions are looked for in the form  
  
\begin{equation}  
R\left( \rho \right) =\exp \left( -\rho ^{2}/4\right) f\,\left( \rho \right).  
\label{11}  
\end{equation}  
\bigskip  
The equation for the function $f\,\left( \rho \right) $ reads  
  
\begin{equation}  
\frac{d^{2}f}{d\rho ^{2}}+\left( \frac{1}{\rho }-\rho \right) \frac{df}{  
d\rho }+\left( \lambda -1-\frac{a}{\rho }\right) f\,\left( \rho \right) =0.  
\label{12}  
\end{equation}  
\bigskip  
For magnetic fields of arbitrary strength  
(the parameter $a$ takes arbitrary values), the regular at the origin  
solution of equation (12), ensuring that $R\left( \rho \right) $  
belongs to the Hilbert space, is given by a series with a complicated  
and almost unknown structure.  
  
It is only for some unique values of the magnetic field that the  
functions $f\,\left( \rho \right) $ are reduced to  
polynomials \cite{1}- \cite{6}, so that the radial functions $R\left( \rho \right) $  
take the form  
  
\begin{equation}  
R_{nk}\left( \rho \right) =C_{nk}\exp \left( -\rho ^{2}/4\right)  
Q_{nk}\,\left( \rho \right) , \label{13}  
\end{equation}  
where  
  
\begin{equation}  
Q_{nk}\,\left( \rho \right) =\sum\limits_{j=0}^{n}b_{j}\rho ^{j}  \label{14}  
\end{equation}  
is the order $n$ polynomial with exactly $k$  zeros ($n$ and $k$ are  
the principle and the radial quantum numbers respectively)  in the  
physical region $\left( \rho \geq 0\right) $. The eigenvalues $\lambda  
$ for all the states of this unique kind are given by the simple  
unified formula  
  
\begin{equation}  
\lambda =n+1,\qquad n+1,2,....  \label{15}  
\end{equation}  
  
Coefficients of the polynomial $Q_{nk}\,\left( \rho \right) $ are  
determined by the recurrence relations  
  
\begin{equation}  
\begin{array}{ll}  
b_{0}= & 1, \\  
b_{1}= & 2, \\  
b_{j}= & \left[ a\;b_{j-1}+\left( j-n-2\right) \;b_{j-2}\right] \;j^{-2}.  
\end{array}  
\label{16}  
\end{equation}  
The unique values of the magnetic fields are determined by the relations  
\begin{equation}  
\left( n+1\right) ^{2}b_{n+1}=ab_{n}-b_{n-1}=0. \label{17}  
\end{equation}  
Several leading values of the parameter follow  
\bigskip  
\begin{equation}  
\begin{array}{ll}  
a_{10}=1, &  \\  
a_{20}=\sqrt{6}, &  \\  
a_{30}=\sqrt{10+\sqrt{73}}, & a_{31}=\sqrt{10-\sqrt{73}}, \\  
a_{40}=\sqrt{25+3\sqrt{33}}, & a_{41}=\sqrt{25-3\sqrt{33}}.  
\end{array} \label{18}  
\end{equation}  
\bigskip  
  
It is of interest to remark that the energy levels for the unique states (13)  
can be obtained from the Bohr - Sommerfeld  quantization rule,  
taking into consideration both the physical  
$\left( \rho \geq 0\right) $ \ and  
nonphysical $\left( \rho <0\right) $ ranges  of the variable $\rho $.  
In this case, the effective potential energy  should be put equal to (see (9))  
  
\begin{equation}  
U_{eff\;}\left( \rho \right) =\frac{1}{4}\rho ^{2}+\frac{a}{\rho }+\frac{1}{  
4\rho ^{2}}.  \label{19}  
\end{equation}  
The Bohr - Sommerfeld quantization rule  with allowance for the  
above remarks can be put into the form  
\bigskip  
\begin{equation}  
2\left( \int\limits_{\rho _{1}}^{\rho _{2}}d\rho \sqrt{\lambda  
-U_{eff\;}\left( \rho \right) }+\int\limits_{\rho _{3}}^{\rho _{4}}d\rho  
\sqrt{\lambda -U_{eff\;}\left( \rho \right) }\right) =2\pi \left( n+1\right).  
\label{20}  
\end{equation}  
\bigskip  
The appearance of unity in the right-hand side of   (20) is explained by  
the fact that there exist four regular lower turning points, each of them  
contributing $1/4.$ It should be mentioned that, when the term $%
1/\left( 4\rho ^{2}\right)$ is not included in expression (19), we have to  
assume that there is an impenetrable potential wall  
at the point $\rho =0$ with the  
contribution $1/2$, instead of $1/4$ \cite{7}. Integration  
in the left-hand side of equation (20) is easily performed with the help of  
the residue technique (the integration contour is depicted in Fig. 3b).  
As a result, we obtain exactly formula (13)  for $\lambda $,  
irrespective of $a$ values.  
  
The physical meaning of the particular magnetic field values  
determined by relations (15) was unclear up to now.  In the following,  
we present considerations that may elucidate the  
physical origin of this phenomenon.  
  
Introduce the center of orbit operators  \cite{8}  
  
\begin{equation}  
\widehat{X}_{c}=-\widehat{p}_{\eta }+\frac{1}{2}\widehat{\xi },\qquad  
\widehat{Y}_{c}=\widehat{p}_{\xi }+\frac{1}{2}\widehat{\eta }.  
\label{21}  
\end{equation}  
Unlike \cite{8}, the symmetric gauge  has been used here,  and  
transformation to the dimensionless form has been performed by means  
of the magnetic length introduced. Then the operator of the square of  
the distance between the center of the orbit and the origin can be  
written in the form  
\begin{equation}  
\widehat{R}_{c}^{2}=\widehat{p}_{\xi }^{2}+\widehat{p}_{\eta  
}^{2}+\frac{1}{4}\widehat{\rho }^{2}+\widehat{L}_{z}.  
\label{22}  
\end{equation}  
In the case of a pure magnetic field, these operator  
commute with the Hamiltonian, so that, in a stationary state, the value  
of $R_{c}^{2}$ is quite well defined. In our case $ R_{c}^{2}$  
can be written in the form  
  
\begin{equation}  
\widehat{R}_{c}^{2}=\widehat{H}-2\widehat{L}_{z}-\frac{a}{\rho }.  
\label{23}  
\end{equation}  
Due to the presence of the Coulomb force,  in our case,  
$R_{c}^{2}$ is not conserved even in a stationary state,  
and hence, only the quantum average of this physical quantity makes  
sense.  
  
Introduce now the average value of the  radius of orbit squared as  
it has been done in \cite{9} for the pure magnetic  
field case  
\begin{equation}  
\left\langle \widehat{R}^{2}\right\rangle =\left\langle \rho  
^{2}\right\rangle -\left\langle \widehat{R}_{c}^{2}\right\rangle .  
\label{24}  
\end{equation}  
With allowance for (4), (15) and for the fact that $l=0$, this equation  
can be written in the form  
  
\begin{equation}  
\left\langle \widehat{R}^{2}\right\rangle =\left\langle \rho  
^{2}\right\rangle +a\left\langle \frac{1}{\rho }\right\rangle -\lambda .  
\label{25}  
\end{equation}  
\bigskip  
Thus, the average magnetic flux in a stationary state (ordinary units  
are again used) takes the form  
  
\begin{equation}  
\Phi =\pi \left\langle \widehat{R}^{2}\right\rangle  
\;B\;l_{B}^{2}=\left\langle \widehat{R}^{2}\right\rangle \,\Phi _{0},  
\label{26}  
\end{equation}  
where $\Phi _{0}=2\pi c\hbar /e$ is the magnetic flux quantum.  
  
Calculations demonstrated that, for stationary states (13), the  
quantity $\Phi $ is the integral multiple  
of $\Phi _{0}$  
  
\begin{equation}  
\Phi =\left( n+1\right) \,\Phi _{0},\quad n=2,3,....  
\label{27}  
\end{equation}  
  
In the following, the results of calculations are illustrated by the  
examples for the cases of states $f_{10}$ and $f_{20}$.  For  the state  
$f_{10}$ we have $a=1,\quad \lambda =2$  
  
\[  
f_{10}\left( \rho \right) =\exp \left( -\rho ^{2}/4\right) \,\left( 1+\rho  
\right) ,  
\]  
  
\[  
\left\langle \rho ^{2}\right\rangle =\left( 10+3\sqrt{2\pi }\right) /\left(  
3+\sqrt{2\pi }\right) ,  
\]  
  
\[  
\left\langle \frac{1}{\rho }\right\rangle =\left( 2+\sqrt{2\pi }\right)  
/\left( 3+\sqrt{2\pi }\right) ,  
\]  
  
\[  
\left\langle \widehat{R}^{2}\right\rangle =2.  
\]  
For the state $f_{20}$ we have $a=\sqrt{6},\quad \lambda =3$  
  
\[  
f_{20}\left( \rho \right) =\exp \left( -\rho ^{2}/4\right) \,\left( 1+\sqrt{6%
}\rho +\rho ^{2}\right) ,  
\]  
\bigskip  
\[  
\left\langle \rho ^{2}\right\rangle =\left( 114+36\sqrt{3\pi }\right) /25+8%
\sqrt{3\pi },  
\]  
\bigskip  
\[  
\left\langle \frac{1}{\rho }\right\rangle =6\left( \sqrt{6}+\sqrt{2\pi }%
\right) /\left( 25+8\sqrt{2\pi }\right) ,  
\]  
\bigskip  
\[  
\left\langle \widehat{R}^{2}\right\rangle =3.  
\]  
\bigskip  
We believe that formula (26) is valid for all the unique  
states, though  we are unable to present a general proof of this  
statement now.  
\bigskip  
  
We are grateful to V.Ch. Zhukovskii and A.V. Borisov for helpful discussions.  
\bigskip  
  

\bigskip  
  
Figure captions  
\bigskip  
  
Figure 1. The curves demonstrate qualitative behaviour of the relation  
$\lambda =2E/\hbar \omega $ as a function of the  parameter  
$a=\sqrt{B_{0}/B}$.  The dots mark the value of  parameters $\lambda $  
and $a$ of unique states.  The lower curve corresponds to the ground  
state, the next one is for the first exited state and so on.  
\bigskip

Fig. 2.  a) The curves of normalized radial  functions of unique states $f _ {nk} \left (r\right) $ in  
physical $ \left (\rho \leq 0\right) $ and nonphysical $ \left (\rho < 0\right) $ ranges at several   
values of principal $ \left (n\right) $ and radial $ \left (k\right) $  quantum numbers.  
  
b) The curves of the normalized radial density of unique states  $D _ {nk} \left (r\right) =r \; f _  
{nk} ^ {2} \; \left (r\right) $ at several  values of principal $ \bigskip \left (n\right) $  
and radial $ \left (k\right) $  quantum numbers.  
  
 The values of a radial variable are measured in the units of the effective Bohr radius $a _ {B} ^  
{\ast} = \hbar ^ {2} /\left (m ^ {\ast} e ^ {2} \right) $.  
\bigskip  
  
Fig. 3.  a) On the curve  of  the  effective potential $U _ {eff} \; \left (\rho \right) $  
in physical $ \left (\rho \leq 0\right) $ and nonphysical $ \left (\rho < 0\right) $  
ranges at $ \lambda =2 $ the turning points  are shown.  
  
       b)   The contour of circumvention of four branching points in the complex plane.  
  
\end{document}